\newcommand{\dm}[2]{(m^2_{#1}-m^2_{#2})}

\documentclass{PoS}

\title{A brief (p)review on a possible fourth generation world to come}

\ShortTitle{4G PReview}

\author{\speaker{George W.S. Hou}
\thanks{Supported by National Science Council and National Center for Theoretical Sciences.}\\
        Department of Physics, National Taiwan University, Taipei, Taiwan 10617\\
        E-mail: \email{wshou@phys.ntu.edu.tw}}

\abstract{..........................\
          ...........................}

\FullConference{35th International Conference of High Energy Physics - ICHEP2010,\\
        July 22-28, 2010\\
        Paris France}

\begin{document}

The $t'$ and $b'$ quarks enjoy nondecoupling in the $Z$-penguin
and box diagrams~\cite{HWS,IL},
bringing in a new $CP$ violating (CPV) phase through
$V_{t's}^*V_{t'b}$~\cite{AH03} into $b\to s$ transitions.
For those who still cite neutrino counting or electroweak
precision tests, please see the ``Four Statements''~\cite{Holdom,
Kribs}.

In the following, we pick up the thread of $B\to K\pi$ direct CPV
(DCPV) difference, linking to the 4th generation prediction (and
current quest) of $t$-dependent CPV (TCPV) in $B_s \to
J/\psi\,\phi$. We then soar up to the heavens with possible source
of CPV for the Universe (CPV-4-U); for the ``cauldron'' that stirs
strong phase transitions, we raise the possible link between
electroweak symmetry breaking and large Yukawa couplings (4-EWSB).
With the prognosis for 2011, 2012 and beyond, we stress these as
preview for the possible 4th generation to come: CPV-4-U and the
\emph{raison d'\^etre} for our Universe (and ourselves); 4-EWSB
and the \emph{raison d'\^etre} for the LHC itself.

\section{Twilight:  $B\to K\pi$ DCPV Difference}

Between BaBar and Belle, DCPV was observed in 2004: $A_{K^+\pi^-}
\cong -10\%$. Though not significant in itself, $A_{K^+\pi^0}$
deviated from $A_{K^+\pi^-}$ by 3.8$\sigma$, which was not
anticipated. In 2008, Belle data alone showed a difference
~\cite{DAKpi} of 4.4$\sigma$: $\Delta A_{K\pi} \cong 16\% >
|A_{K^+\pi^-}|$ is strikingly large!

Struck by the strength of the difference, we utilized
nondecoupling and the new complex factor of $V_{t's}^*V_{t'b}$ and
demonstrated~\cite{HNS}, using PQCD factorization at LO
(equivalent to QCDF at NLO), that the $t'$ quark can in principle
generate $\Delta A_{K\pi}$. The work was further
refined~\cite{HLMN} at NLO in PQCD, where having some
``color-suppressed tree'' amplitude $C$ made it easier to account
for $\Delta A_{K\pi}$. As the $b\to s$ $Z$-penguin and $b\bar s
\leftrightarrow \bar b s$ box diagram are cousins, we
predicted~\cite{HNS} $\sin2\Phi_{B_s}$ (CPV phase of the said box
amplitude) to be in the range of $-0.2$ to $-0.7$. It was further
illustrated~\cite{HNS05} how the complete $4\times 4$ CKM matrix
could be determined, by facing all flavor constraints.

Boxes and $Z$-penguins have provided us with a wealth of
information on flavor physics and CPV. From the $K^0$--$\bar K^0$
box, we learned the GIM mechanism and existence of charm, and
accounted for $\varepsilon_K$; we inferred heavy top from the
$B_d$ box, and eventually measured $\sin2\phi_1/\beta$. With heavy
top, the $s\to d$ $Z$-penguin diminished the value for
$\varepsilon'/\varepsilon$ and gave the value for $K\to
\pi\nu\bar\nu$, giving $Z$-penguin dominance of $b\to
s\ell^+\ell^-$ and the measurable top effect in $Z\to b\bar b$.
All that, with 3 generations. Just wait if there exists a 4th
generation --- it will bring out an agenda for all aspects of
flavor physics and CPV: $B_s$, $A_{\rm FB}(B\to K^*\ell^+\ell^-)$,
renewed interest in $Z\to b\bar b$, $\varepsilon_K$,
$\sin2\phi_1/\beta$, and to a lesser extent,
$\varepsilon'/\varepsilon$ and $D$ mixing. But as we shall see,
the most exciting aspect would be the direct search for $t'$ and
$b'$ quark at the LHC, with continued interest at the Tevatron.

\vskip0.15cm \noindent{\textbf{\underline{PAMELA!}}}

But let me bring myself back to how things were before 2008, by
drawing analogy with ``PAMELA''. The excess of energetic $e^+$
observed by PAMELA and others could in fact be due to nearby
pulsars~\cite{PAMELA}, hence astrophysical. BUT THAT DIDN'T STOP
THE DM (Dark Matter) PARTICLE SPECULATOR/THEORISTS. In contrast,
as alluded to, the $B\to K\pi$ DCPV difference could also arise
from ``enhanced color-suppressed $C$'', AND THIS SEEMS TO STOP
FURTHER THOUGHTS ACROSS (the) ATLANTIC!? The possibility of new
physics in $b\to s$ $Z$-penguin, and the prediction for large CPV
effect in $B_s \to J/\psi\,\phi$, was largely ignored during
2005-2007.
We remark that QCDF did not predict $A_{K^+\pi^-}$, while SCET got
the sign for $\Delta A_{K\pi}$ totally wrong, \emph{after taking
$A_{K^+\pi^-}$ as input}. It was PQCD which anticipated, ca. 2001,
the sign as well as strength of $A_{K^+\pi^-}$. As much as
experimentalists do ``blind analysis'', theorists should make
``predictions''.

\section{Moonshine:  Prediction/Quest for TCPV in $B_s \to J/\psi\,\phi$}

In the CDF public note CDF/ANAL/BOTTOM/PUBLIC/9458, it is stated
``George Hou predicted the presence of a $t'$ quark \ldots to
explain the Belle result and predicted \emph{a priori} the
observation of a large $CP$-violating phase in $B_s \to
J/\psi\,\phi$ decays [7, 8]''. ``Ref. [7]'' is our
Ref.~\cite{HNS}, while ``Ref.~[8]'' is a refinement~\cite{HNS07}
of $\sin2\Phi_{B_s}$ prediction to the range of $-0.5$ to $-0.7$
(SM value is $\sim -0.04$). Surprisingly, three measurements, by
CDF in 12/2007 and 8/2008 and D$\emptyset$ in 2/2008, gave central
values right in this range!! The error was, of course, large:
despite some rattling by UT\emph{fit}, and taking a year to
combine the two experiments, the significance was only
2.1$\sigma$~\cite{Punzi} in summer 2009. Note that CDF uses
$-\sin2\beta_s$ while D$\emptyset$ uses $\sin\phi_s$ instead of
our $\sin2\Phi_{B_s}$.

Summer 2010 was volatile. First, D$\emptyset$ splashed forth a
value of $A_{\rm SL}$ that deviated by 3.2$\sigma$ from SM
expectations.\footnote{ %
We remark that the D$\emptyset$ value of $A_{\rm SL}$ violates a
bound. If it is confirmed in the future, e.g. by LHCb, it probably
implies hadronically enhanced $\Delta \Gamma_s$ from OPE
predictions. See~\cite{HouTOP10}; we do not believe in New Physics
enhanced $\Delta \Gamma_s$.} A week later, CDF uncovered their new
$\sin2\beta_s$ value, which dropped to only $\sim 1\sigma$ away
from SM. Even though D$\emptyset$ released a larger new value for
$\sin\phi_s$, the verdict seems to be that $\sin2\Phi_{B_s} \equiv
\sin\phi_s \equiv -\sin2\beta_s$ is weaker in strength than in
2009. Interestingly, we had just studied the case for heavier
$m_{t'} =$ 500 GeV (rather than 300 GeV in earlier studies),
turning out a central value of $\sin2\Phi_{B_s} \sim
-0.33$~\cite{HM10}. A heavier $t'$ seems preferred now by CPV
data.

The sad thing for Tevatron is, with a target central value of
$-0.3$ rather than $-0.6$, there is little hope that the combined
Tevatron result would ever reach ``evidence'' level, given that
CDF and D$\emptyset$ results are already using datasets of order
5--6 fb$^{-1}$. But it is assured that the value should be quickly
measured by LHCb, once it has even a couple of 100 pb$^{-1}$,
which could arrive before summer 2011. Another thing worthy of
note is that the 4th generation seems ``rehabilitated'' since
start of 2010, judging from significant work even just on flavor
and CPV physics~\cite{4G10}.

\section{Starry Heavens:  CPV 4 Universe ?}

Recall the Sakharov conditions for generating Baryon Asymmetry of
the Universe (BAU): Baryon Number Violation;  CP Violation;  Out
of Equilibrium. In his Nobel Lecture, Kobayashi admitted that
``Matter dominance of the Universe seems requiring new source of
CP violation'', beyond his model with Maskawa.
We illustrate this with a simple dimensional analysis of the
Jarlskog invariant for CPV, $J \equiv {\rm
Im}\,\det\left[m_um_u^\dag,\ m_dm_d^\dag\right]$ for 3
generations. Expanded, one has
\begin{equation}
J = \dm{t}{u} \dm{t}{c} \dm{c}{u}
    \dm{b}{d} \dm{b}{s} \dm{s}{d} \, A,
 \label{eq:JinSM3}
\end{equation}
where $A \simeq 3\times 10^{-5}$ is the common triangle area, e.g.
for $V_{ud}^*V_{ub} + V_{cd}^*V_{cb} + V_{td}^*V_{tb} = 0$. One
has CPV \emph{if and only if} $J \neq 0$. Normalizing by
electroweak phase transition temperature $T \simeq 100$ TeV, one
finds $J$ seems short by $10^{-10}$ (if not more) to account for
BAU.

As a byproduct of ``\emph{Nature} writing''~\cite{DAKpi}
(explaining CPV ``to biologists"), I noticed that if one had 4
generations, then simply shifting ``123'' in Eq.~(\ref{eq:JinSM3})
to ``234'', one finds~\cite{Hou09},
\begin{equation}
J_{(2,3,4)}^{sb} = \dm{t'}{c} \dm{t'}{t} \dm{t}{c}
           \dm{b'}{s} \dm{b'}{b} \dm{b}{s} A_{234}^{sb},
 \label{eq:J234_Hou}
\end{equation}
which would be enhanced by $\sim$ 15 orders of magnitude over $J$
of Eq.~(\ref{eq:JinSM3}), brought about by the heavy $t'$, $b'$
quark masses, which are taken to be in the range of 300 to 600
GeV. The CPV area factor $A_{234}$ (found phenomenologically to be
${\cal O}(10-30)$) should not be too different from 1.
A detailed algebraic check showed~\cite{HMS10} that indeed
Eq.~(\ref{eq:J234_Hou}) is the leading effect, which is due to
mass hierarchy mostly. Subleading effects, which are many, are
typically at $\sim 1/10$. Note that this 1000 trillion enhancement
comes from large Yukawa couplings, hence is dynamical.

\section{Cauldron:  Large Yukawa and EWSB ?}

The $t'$ and $b'$ mass bounds are becoming rather heavy at the
Tevatron, implying that we are entering the realm of large Yukawa
couplings for the search of new heavy chiral fermions.

CDF has pursued $t' \to Wq$ search ($q$ jet flavor unspecified)
for several years, now joined by D$\emptyset$ this summer,
with~\cite{Lister} $m_{t'} < 335$ and 296 GeV excluded at 95\%
C.L., respectively. However, for both experiments, the observed
bounds are somewhat weaker than expected sensitivities. In fact,
both experiments show an inkling of excess ($\sim 2\sigma$) events
at high $H_T$ and reconstructed mass. It is not clear whether this
is genuine, or common underestimation of QCD backgrounds.
CDF has also pursued $b' \to Wt$ search~\cite{Scodellaro}, where
the clean signature of same-sign dileptons give rise to a bound of
338 GeV. A more stringent bound is inferred from leptons plus
(multi-)jets, but it may be questionable in this case how well one
understands QCD background with high number of jets.

In any case, the bounds are now above 300 GeV, and one is wary of
the so-called ``unitarity bound'': partial wave unitarity breaks
down~\cite{CFH79} starting 500-600 GeV! Of course, probability is
always conserved, so the breakdown only reflects the fact that
strong Yukawa couplings set in, and perturbation theory breaks
down. This reminds one further of an old suggestion by Nambu:
Could EWSB be due to heavy chiral quarks, $t'$ and $b'$, above the
unitarity bound? I.e., the conjecture that EWSB might be due to
$\bar QQ$ condensation ($\langle\bar QQ\rangle \neq 0$) induced by
large Yukawa couplings. To seriously address this, one would need
to study Higgs-Yukawa sector on a lattice.

We offer some speculation, keeping an ``experimentalist'' mindset.

The SU(2)$_L\times$U(1) \emph{chiral} gauge symmetry is
experimentally established. Spontaneous symmetry breaking is also
experimentally established, with massive $W$ and $Z$ bosons, and
massive fermions, too. Since renormalizability (needed for contact
with LEP data) depends only on Ward identities, it is unaffected
by SSB. One can then take the physical gauge, where no would-be
Goldstone bosons, or scalar particles, appear at all.
Consider the $gV_{ij}\,\bar u_i \gamma_\mu L d_j\, W^\mu$ gauge
coupling vertex, which involves only left-handed quarks. Replacing
the $W^\mu$ by $k^\mu/M_W \equiv \hat{k}^\mu$ from the
longitudinal $W$ propagator, and contract with $\gamma_\mu$.
Noting that $k = p_i - p_j$, one finds
\begin{equation}
 g\,\hat{k}{\hskip-0.17cm}/\,L
  = g\,(\hat{p}{\hskip-0.17cm}/_i - \hat{p}{\hskip-0.17cm}/_j)\,L
  = g\,(\hat{m}_i\,L - \hat{m}_j\,R) = \lambda_i\,L - \lambda_j\,R,
\end{equation}
where we use the equation of motion to replace momenta by mass,
and in so doing, generated the chirality flip. In the last step,
we produce the ``Yukawa coupling'' of the Goldstone boson --- but
we started from the gauge coupling! This is not surprising, as the
Goldstone particle couples to mass. However, notice we never
mentioned the existence of the Higgs particle in our entire
discussion.

If a new heavy chiral quark $Q$ exists, can its effective Yukawa
coupling generate $\langle\bar QQ\rangle \neq 0$?

\section{Prognosis --- 2011, 2012, beyond}

On the $\sin2\Phi_{B_s} \equiv \sin\phi_s \equiv -\sin2\beta_s$
front, there will be a tug-of-war between Tevatron and LHCb,
depending on its value and how fast LHCb can deliver. But no doubt
LHCb would end up the winner. By the way, \emph{I will put 4G on
the back-burner if $\sin2\Phi_{B_s}$ turns out to be SM3-like}.

As for $t'/b'$ direct search, things would also turn towards LHC's
advantage, once some data is accumulated: bounds would surpass
Tevatron's with just 100 pb$^{-1}$~\cite{CMS7TeV}. With 1
fb${^{-1}}$, the target for 2011, the expected bound would reach
the unitarity bound! Thus, to continue the pursuit beyond 2012,
one would need guidance from lattice Higgs-Yukawa studies.
We remark that, in the long run (beyond 2020), one could in
principle extract CPV phase via the $b' \to s\gamma$
mode~\cite{AH09}.

If the 4th generation is discovered at the LHC, one important
measurement to fix the flavor/CP parameters would be~\cite{HNS05}
$K_L \to \pi^0\nu\bar\nu$, which could be done by the KOTO
experiment at J-PARC. Though a plethora of measurements can be
made, it is not yet clear whether the Super B factories, such as
Belle II, could provide definitive measurements. This is a
question to be pursued.

\end{document}